\documentclass[prl,twocolumn,superscriptaddress,showpacs]{revtex4}
\usepackage{amsmath}
\usepackage{amssymb}
\usepackage{graphicx}

\begin{document}
\title{Dispersive Excitations in the High-Temperature 
Superconductor La$_{2-x}$Sr$_x$CuO$_4$}

\author{N.B. Christensen}
\affiliation{Materials Research Department, Ris\o\ National 
Laboratory, Denmark}
\author{D.F. McMorrow}
\affiliation{Materials Research Department, Ris\o\ National Laboratory, Denmark}
\affiliation{London Centre for Nanotechnology and Department of Physics and Astronomy,
University College London, London, UK}
\author{H.M. R\o nnow}
\affiliation{ETH-Z\"{u}rich \& Paul Scherrer Institut, Switzerland}
\affiliation{James Franck Institute, University of Chicago, USA}
\author{B. Lake}
\affiliation{Clarendon Laboratory, University of Oxford, UK}
\author{S.M. Hayden}
\affiliation{H.H. Wills Physics Laboratory, University of Bristol, UK}
\author{G. Aeppli}
\affiliation{London Centre for Nanotechnology and Department of Physics and Astronomy,
University College London, London, UK}
\author{T.G. Perring}
\affiliation{ISIS Facility, Rutherford Appleton Laboratory, UK}
\author{M. Mangkorntong}
\affiliation{Institute of Solid State Physics, University of Tokyo, Japan}
\author{M. Nohara}
\affiliation{Institute of Solid State Physics, University of Tokyo, Japan}
\author{H. Takagi}
\affiliation{Institute of Solid State Physics, University of Tokyo, Japan}

\begin{abstract}
High-resolution neutron scattering experiments on optimally doped La$_{2-x}$Sr$_x$CuO$_4$ ($x$=0.16)
reveal that the magnetic excitations are dispersive.
The dispersion is the same as in YBa$_2$Cu$_3$O$_{6.85}$, and is quantitatively related to
that observed with charge sensitive probes. 
The associated velocity in La$_{2-x}$Sr$_x$CuO$_4$ appears to be weakly dependent on doping
with a value close to the spin wave velocity of the insulating ($x$=0) parent compound.
In contrast to the insulator, the excitations broaden rapidly with
increasing energy, forming a continuum at higher energy, and  bear 
a remarkable resemblance to multi-particle excitations observed in 1D S=1/2 antiferromagnets. 
The magnetic
correlations are 2D, and so  rule out the simplest scenarios where the copper oxide planes
are subdivided into weakly interacting 1D magnets.   
\end{abstract}

\maketitle

The defining properties of any particle or wave-like excitation are the dimensionality of the
space in which it moves and its dispersion.
Conventional magnets host spin waves, which are
coherently precessing deviations of the spins from their equilibrium, ordered
configuration. The antiferromagnetic (AF), insulating parents of the
high-temperature, CuO$_2$ based superconductors are no exception to this rule \cite{Hayden1991}. An
important question is what happens to the magnetic excitations upon chemical doping to
produce metals and superconductors. Here we describe experiments
on La$_{2-x}$Sr$_x$CuO$_4$ (LSCO), which reveal that the magnetic excitations are in
fact dispersive and appear, at least at low energies, to be two-dimensional (2D).
However, they are not derived from propagating
excitations in the manner of ordinary spin waves. We show how these
dispersive spin excitations bear a quantitative relation to those in the YBa$_2$Cu$_3$O$_{6+x}$
(YBCO) family \cite{Dai2001,Bourges2000}, and more astonishingly, to charge
sensitive measures of electronic excitations \cite{Hoffman2002,McElroy2003}. This discovery clearly establishes the
notion of universality amongst the different cuprate families, and will greatly simplify
the theory for this highly varied class of complex materials.

The technique chosen is inelastic neutron scattering, which
measures the spin excitation spectrum,
$\chi^{\prime\prime}({\mathbf Q},\omega)$, as a function of both momentum $\hbar{\mathbf Q}$
and energy $\hbar\omega$. Many neutron
experiments have been performed on the superconducting (SC) cuprates, and it
is fair to ask why another is needed. The reason is that the new MAPS
spectrometer at ISIS, UK finally can provide the long sought for
definitive results by offering global momentum and energy surveys
with momentum resolution as small as 0.02 \AA$^{-1}$ rather than the more typical 0.1 \AA$^{-1}$.
Nine single crystals of La$_{1.84}$Sr$_{0.16}$CuO$_4$  (total mass of 18.7 g, $T_c=$38.5 K) grown by the floating
zone method were co-aligned with a resulting mosaic spread of 1.5$^\circ$. The sample was
mounted with the SC planes perpendicular to the beam of neutrons with
incident energy of 55 meV. Data were collected at 10 and 40 K, and were converted to
$\chi^{\prime\prime}({\mathbf Q},\omega)$ in absolute units of $\mu_B^2$ eV$^{-1}$
per formula unit (f.u.).

\begin{figure}
\includegraphics[scale=0.49,bb=29 269 530 576,clip]{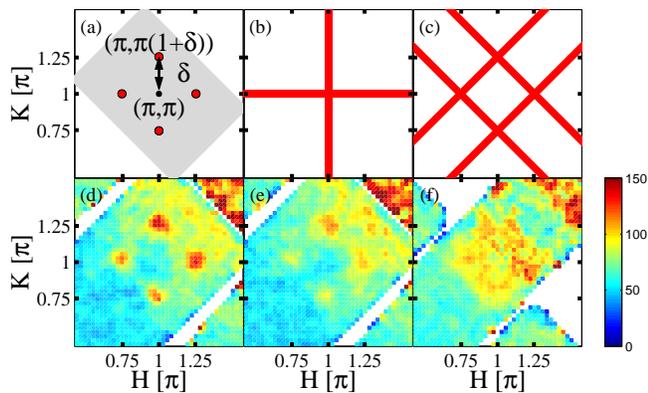}
\caption{(a)-(c) 2D reciprocal space indicating the
diffraction patterns for different scenarios of the magnetic correlations.
(d)-(f) Images from the MAPS spectrometer of the
dynamical magnetic susceptibility $\chi^{\prime\prime}({\mathbf Q},\omega)$
for LSCO $x=0.16$.
Colour scale: intensity in units of $\mu_B^2$ eV$^{-1}$ f.u.$^{-1}$. (d) and (f),
$T=$10 K in the SC phase for  $\hbar\omega=$10 and 30 meV respectively. (e) $T=$40 K in
the normal phase at 10 meV.
}
\label{fig1_prl}
\end{figure}

The undoped parent La$_2$CuO$_4$ displays commensurate AF order giving
diffraction at ${\mathbf Q}=(\pi,\pi)$ (Fig.\ \ref{fig1_prl}(a)).
Upon doping, the diffraction peak splits into four incommensurate (IC)
spots \cite{Yamada1998,Lake1999} reflecting the fact that in real space the AF correlations have
become modulated with a period inversely related to $\delta\pi$, the distance in reciprocal space
from $(\pi,\pi)$.
In one very popular model \cite{Tranquada1995,Zaanen2001,Kivelson2003}, the modulation is due to a tendency of the
CuO$_2$ planes to break into stripes containing 1D magnets.
If these magnets were truly independent of each other, they would give rise to 
perpendicular streaks of scattering.
Since there are
two equivalent (orthogonal) directions to choose from, a typical sample
would produce orthogonal streaks passing through $(\pi,\pi)$ (Fig. \ref{fig1_prl}(b)).
Only an internal modulation along each stripe, as occurs for spin $1/2$ chains
doped with holes or in an external magnetic field \cite{Tsvelik}, can
produce IC features in the neutron scattering images, such as the crossing
of streaks from diagonal stripes as depicted in Fig. \ref{fig1_prl}(c).

Figure \ref{fig1_prl}(d)-(f) displays images of $\chi^{\prime\prime}({\mathbf Q},\omega)$
for La$_{1.84}$Sr$_{0.16}$CuO$_4$, produced directly from the
raw data without performing any background subtraction. They are the first to afford
complete (high) isotropic resolution coverage of reciprocal space and energy for
superconducting LSCO, and directly reveal a very important fact. In the
SC phase at $T=$10 K and at low energies (Fig. \ref{fig1_prl}(d), $\hbar\omega=$10 meV) the
magnetic response starts out as sharply isotropic peaks at the quartet of IC
positions schematically illustrated in Fig. \ref{fig1_prl}(a), with essentially no scattering at ${\mathbf Q}=(\pi,\pi)$
and no evidence for the streaks sketched in Fig. \ref{fig1_prl}(b) and \ref{fig1_prl}(c). The isotropy of the
peaks and absence of obvious streaks implies \cite{Zaanen2001,Kivelson2003}  that the magnetic correlations are
2D and immediately rules out any scenario 
based on a naive subdivision of
the material into weakly coupled magnetic stripes. Until now, the extent of this isotropy
has not been clear because  data have been collected using
instruments whose resolution functions are highly anisotropic. 
If a stripe picture is used to interpret our data, 
there must be strong inter-stripe as well as intra-stripe exchange coupling.


\begin{figure}
\includegraphics[scale=0.47,bb=38 186 520 575,clip]{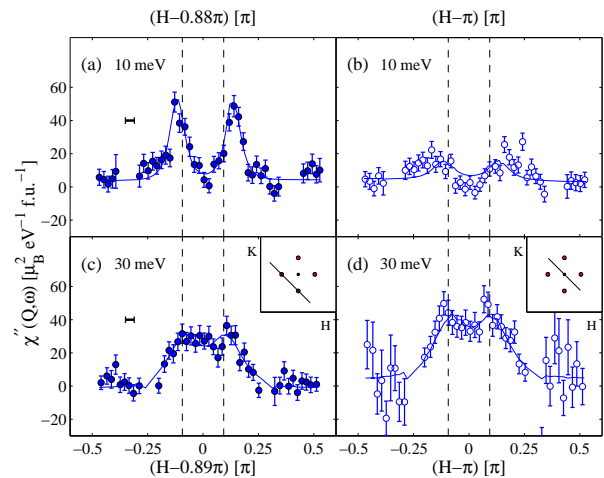}
\caption{Constant energy cuts from MAPS through $\chi^{\prime\prime}({\mathbf Q},\omega)$
for La$_{1.84}$Sr$_{0.16}$CuO$_4$ at $T=$10 K. The scan trajectory is transverse as
shown in the insets to (c) and (d). (a)
and (b): $\hbar\omega=$10 meV; (c) and (d): $\hbar\omega=$30 meV.
Solid lines: fits to the model discussed in the text. Vertical
dashed lines: fitted peak positions at 30 meV. Horizontal bars in (a) and
(c) represent the instrumental resolution (FWHM). Using the wide reciprocal space coverage of MAPS it was possible to 
determine accurately the background at positions away from the magnetic scattering. This background has been 
subtracted from the data shown here. 
}
\label{fig2_prl}
\end{figure}

Warming to just above $T_c$ (Fig. \ref{fig1_prl}(e)), the IC peaks
weaken, as spectral weight is shifted to fill
the spin gap at lower energy \cite{Lake1999}. We gain dramatic new knowledge by going to 30 meV
(Fig. \ref{fig1_prl}(f)), where we see that the peaks are no longer well-defined. Instead,
the scattering appears more as a ``box" surrounding $(\pi,\pi)$, and there is now also
substantial scattering at $(\pi,\pi)$.

In Fig. \ref{fig2_prl}, we examine the data in more detail by looking at constant-energy cuts
through the images in Fig. \ref{fig1_prl}. At 10 meV,
Fig. \ref{fig2_prl}(a) displays the  IC
peaks  at low energies, while in Fig. \ref{fig2_prl}(b) we see very weak sides to the
IC ``box". 
By increasing $E$ to 30 meV it is clear that the response has moved in
towards $(\pi,\pi)$ (Fig. \ref{fig1_prl}), meaning that it is dispersive. In addition, the peaks lose definition $-$ the
corners (Fig. \ref{fig2_prl}(c)) are now no sharper than the walls (Fig. \ref{fig2_prl}(d)) of the ``box", and both have
broadened so that there is considerable scattering even at $(\pi,\pi)$. This transition to
broader, continuum like scattering implies that we are not dealing with well-defined
bosonic modes of a quantum spin liquid or AF. A key result of our work, namely that the IC
peaks disperse and cannot be due to conventional propagating modes, is thus seen by
straightforward display of the data.

\begin{figure}
\includegraphics[scale=0.5,bb=112 222 455 572,clip]{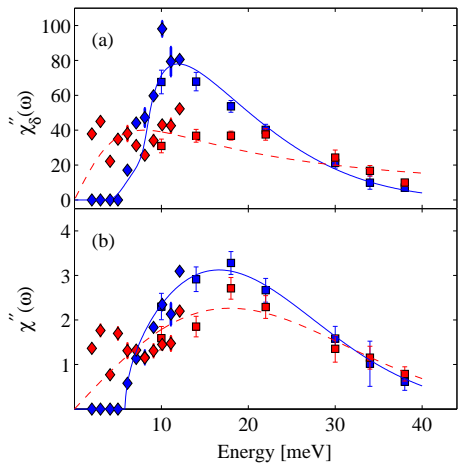}
\caption{ Energy dependence of $\chi^{\prime\prime}({\mathbf Q},\omega)$
for optimally doped LSCO in units of $\mu_B^2$ eV$^{-1}$ f.u.$^{-1}$.
(a) Fitted peak amplitude at ${\mathbf Q}_\delta$: Superconducting phase, blue symbols; normal
phase, red symbols. MAPS data, (squares); triple-axis data (diamonds). (b): $\chi^{\prime\prime}(\omega)$
symbols as in (a). Lines are guides to the eye. The region 24-28 meV was found to be
strongly contaminated by phonon scattering and was not analysed.}
\label{fig3_prl}
\end{figure}

\begin{figure}
\includegraphics[scale=0.45,bb=27 197 589 582,clip]{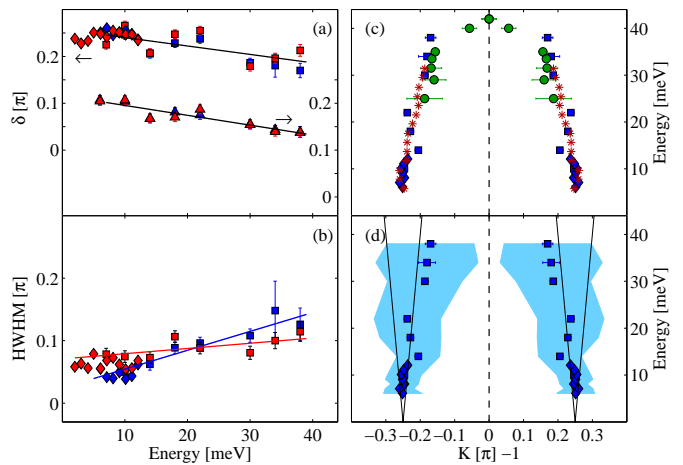}
\caption{(a) Incommensurability $\delta$ in LSCO. Squares: MAPS data for optimal doping, $x$=0.16. 
Triangles: MAPS data for LSCO $x$=0.10. Diamonds:  triple-axis data on LSCO $x$=0.16. 
Blue symbols, superconducting phase; red symbols, normal state.  (b)
Energy dependence of the half width at half maximum. Symbols as in (a). (c) Comparison
of $\delta$ for the superconducting phases of optimally doped LSCO (symbols as in (a))
and YBCO$_{6.85}$ (circles,  \cite{Ronnow2000}), and half the wavevector of the electronic excitations (stars,
 \cite{Hoffman2002}) observed along the $(1,0)$ direction in Bi$_2$Sr$_2$CaCu$_2$O$_{8+\delta}$ using STM. (d)
Dispersion of the excitations in LSCO. Shaded regions represent the fitted FWHM. 
Solid lines show the spin-wave velocity of undoped La$_2$CuO$_4$  with  $J=$156 meV \cite{Hayden1991}. 
The images in (c) and (d)
have been symmetrised for display.}
\label{fig4_prl}
\end{figure}

We analysed the momentum-dependent images using
\begin{equation}
\chi^{\prime\prime}({\mathbf Q},\omega)=\sum_{\delta}\frac{\chi_\delta\,\kappa^4(\omega)}
{(\kappa^2(\omega)+({\mathbf Q}-{\mathbf Q}_\delta(\omega))^2)^2},
\end{equation}
where $\kappa(\omega)$ is an inverse correlation length, and the sum runs over the four IC positions.
A good description of the data was obtained by convoluting the above lineshape with
the full instrumental resolution function and fitting it to the data (see Fig. \ref{fig2_prl}).
From this analysis values were obtained of the
incommensurability $\delta$, inverse correlation length $\kappa(\omega)$, and intensity $\chi_\delta$ of the magnetic
fluctuations. 
The values of $\delta$ are plotted in Fig. \ref{fig4_prl}(a). It is as apparent from the fits as 
from the raw data that the peaks in the response
move closer to ${\mathbf Q}=(\pi,\pi)$ and broaden with increasing energy. No change 
of $\delta$ is observed on entering the
SC state. To emphasize that dispersion of the IC mode is a
generic feature of LSCO we also show data taken using MAPS for an underdoped
LSCO sample, $x$=0.10, where the initial incommensurability is lower, but the velocity
is the same. What is even more remarkable is that the velocity associated with dispersion 
of the non-propagating spin excitations seen here for the metals is the same ($\sim$ 1eV~\AA) 
as that for the propagating spin waves in the parent insulator \cite{Hayden1991}.  
Figure \ref{fig4_prl}(b) shows the peak width, which above 20 meV  is
indistinguishable in the normal and SC phases; at lower energies the excitations become more
coherent in the SC state\cite{Lake1999}.

Figure \ref{fig3_prl}(a) displays peak intensities,
including data from the same crystals obtained at
the RITA spectrometer, Ris\o\ , Denmark. As observed
previously, superconductivity redistributes the spectral weight from energies below the
spin gap $\Delta=$7 meV, to higher energies \cite{Lake1999}. What is new is that we are able now to see
the full extent of the redistribution. The net effect is to produce a peak in
$\chi^{\prime\prime}({\mathbf Q},\omega)$
centred at 12$\pm$2 meV, although the redistribution takes place for energies up to about 30
meV. Figure \ref{fig3_prl}(b)
presents the local susceptibility $\chi^{\prime\prime}(\omega)$
calculated by integrating the response over the 2D Brillouin zone. In the
SC phase,
$\chi^{\prime\prime}(\omega)$ is peaked at 18$\pm$2 meV, with a half width at half
maximum of 12$\pm$2 meV. The normal state mean-squared fluctuating moment \cite{Dai1999},
calculated from $\chi^{\prime\prime}(\omega)$
up to 40 meV is $\langle m^2\rangle=$ 0.062$\pm$0.005 $\mu_B^2$(CuO$_2$)$^{-1}$
and is unchanged on cooling through $T_c$.
In other words, the spectral weight removed by the
opening of the spin gap is preserved and merely shifted to higher energy.
A similar shift of spectral weight leads to the formation of a commensurate resonance in
lightly doped YBa$_2$Cu$_3$O$_{6+x}$ \cite{Dai1999,Stock2004}.
In the case of LSCO, the amount of spectral weight shifted is
$\delta\langle m^2\rangle=$0.010$\pm$0.005 $\mu_B^2$(CuO$_2$)$^{-1}$,
of the same order as $\delta\langle m^2\rangle=$0.03 $\mu_B^2$(CuO$_2$)$^{-1}$,
in the resonance of YBa$_2$Cu$_3$O$_{6.6}$ \cite{Dai1999}. That the spectral weight in LSCO is conserved
already up to 40 meV, makes the existence of the long sought after higher frequency
commensurate ``resonance" at ${\mathbf Q}=(\pi,\pi)$ as in YBCO \cite{RossatMignod1991,Fong1997,Mook1998} very improbable.

It has often been stated that the IC fluctuations in
LSCO are dispersion-less 
and a pathology of this single layer material \cite{Bourges2000,Yamada1998}.
We argue here that this view is incorrect, even though $-$
at first glance $-$ YBCO, whose fundamental building
blocks are CuO$_2$-bilayers, displays a quite different magnetic response. Below $T_c$,
$\chi^{\prime\prime}({\mathbf Q},\omega)$
is dominated by a ``resonance" peak at ${\mathbf Q}=(\pi,\pi)$ ($E_{\rm res}=$41 meV in the case of
optimal doping) \cite{RossatMignod1991,Fong1997,Mook1998}.
Subsequent work revealed that at lower energies, $E < E_{\rm res}$, the
response becomes IC \cite{Dai2001,Bourges2000,Ronnow2000}. Here we have discovered that the
IC response in LSCO is dispersive and remarkably similar to that of
YBCO$_{6.85}$ as shown in Fig. \ref{fig4_prl}(c).
The main difference between the two systems is that the spectral weight redistribution in LSCO takes
place  well below the energy  where the IC modes begin to merge (Fig. \ref{fig4_prl}(d)).
In YBCO the larger spin gap results in  spectral weight being pushed  into
the region where the IC modes merge, forming a sharp peak-like
response.

While our data resemble those for YBCO, they differ from those 
for conventional magnets, even with long-period order. For example, La$_{1.69}$Sr$_{0.31}$NiO$_4$ 
\cite{Bourges2003}, displays two sharply defined modes 
emanating symmetrically from each IC wavevector as indicated by the solid lines in Fig. \ref{fig4_prl}(d). 
More recently a study \cite{Tranquada2004} of non-superconducting
La$_{1.875}$Ba$_{0.125}$CuO$_4$, reported asymmetric dispersion but did not specifically address the 
lineshape and hence provided no information on any broadening at low energies.
In contrast, our high-resolution experiments show that the 
magnetic excitations disperse in only one direction (inwards), 
and broaden with increasing energy as rapidly as they disperse (Fig. \ref{fig4_prl}(d)). This 
means that we are not dealing with propagating spin-wave modes of the type found in 
conventional magnets. 

Instead, our observed response is best described as resulting from a
continuum of excitations, indicating
that the neutron does not excite the fundamental quasiparticles
of the system, but rather coherent pairs, quartets and so on, 
of quasiparticles. There are several candidates for the underlying  quasiparticles. 
In the ordinary theory of metals the underlying particles are simply electrons and
the multi-particle excitations, which neutron scattering would observe are correlated
electron-hole pairs.  Many theories attempt to explain cuprate spin fluctuations
in terms of propagating electrons and holes described by an underlying band
structure\cite{Kao2000,Norman2000,Chubukov2001}.  However, we lack a single
model which describes the remarkable robustness of the velocity with respect to doping
(Fig. \ref{fig4_prl}(a)), the absence of measurable dispersion in the spin gap \cite{Lake1999},  and the universality  
of the observed response between the bilayer and single layer materials (Fig. \ref{fig4_prl}(c)).
A second possibility for the underlying excitations is suggested by returning
to the concept of spin-charge separation introduced to the cuprates almost
immediately after the discovery
of high-$T_c$ \cite{Anderson1987,Rokhsar1988} superconductivity.
Here the fundamental quasiparticles are no longer
electrons, carrying spin $1/2$ and charge $-e$; instead, they are charge neutral objects,
spinons carrying spin $1/2$, and spin-less objects, holons, carrying charge $+e$.
While this picture has been verified in 1D \cite{Tsvelik,Tennant1993}, its
applicability in 2D has been more controversial. Spin-charge separation in the cuprates could occur
by sub-division into 1D chains \cite{Dagotto1996}, as suggested by the simplest
stripe picture \cite{Zaanen2001,Kivelson2003}, or via the effect of competing interactions
on 2D quantum systems \cite{Coldea2003,Lee1992}. 
Further work is required to differentiate between these two possibilities.
Both descriptions imply proximity to quantum critical points, 
which naturally lead to a magnetic response dominated by non-propagating modes, and has given 
a very natural framework for understanding the low-frequency properties of an LSCO sample with $x$=0.14 \cite{aeppli1997}, 
close to the special filling fraction $x$=1/8.  

We end by pointing out a correlation between our data and recent STM experiments which have observed 
dispersing charge quasiparticles. Specifically, the stars in Fig.\ref{fig4_prl}(c) 
represent the position along [1,0] of the most obvious IC peak from STM data (positive bias) 
for ``as grown" Bi$_2$Sr$_2$CaCu$_2$O$_{8+\delta}$ \cite{Hoffman2002}. 
Within the conventional band theory of ordinary paramagnetic metals, 
the loci of the charge and spin responses in Q-$E$ space should coincide. 
This is actually not the case $-$ the STM wavevectors when divided by a factor of two 
line up with the neutron data. 
This factor of two is of course familiar from the static order seen in conventional magnets
\cite{Tranquada1995,zachar1998}. 
It is quite astonishing that, once this factor of two is included,  
the spin and electronic response functions measured by neutrons and 
STM for different materials display nearly identical
dispersion.
At present the question is open as to 
whether this correlation between the neutron and STM data sets is 
coincidental or whether the two are profoundly connected. 

We acknowledge stimulating discussions with B. M{\o}ller
Andersen, S. Davis, S. Chakravarty, T. Giammarchi, P. Hedeg\aa rd, S. Kivelson, and
J. Mesot, and are especially grateful to T. Rosenbaum for his interest and support from the 
University of Chicago through the National Science
Foundation Grant No. DMR-0114798. Work in London was supported by a
Wolfson-Royal Society Research Merit Award and the Basic Technology program of
the UK research councils.


\end{document}